\documentclass[sigconf, nonacm]{acmart}

\usepackage{xspace}
\usepackage{xcolor}
\usepackage{color}
\usepackage{colortbl}
\usepackage{makecell}
\usepackage{multirow}
\usepackage{geometry}
\usepackage{hhline}
\usepackage{marginnote}
\usepackage{hyperref}
\usepackage[normalem]{ulem}
\usepackage{pifont}
\usepackage{subcaption}

\usepackage[lined,boxed,vlined,ruled,linesnumbered]{algorithm2e}

\newcommand{\zw}[1]{\textcolor{purple}{#1}}
\newcommand{\zxh}[1]{\textcolor{red}{#1}}

\newcommand{\blue}[1]{\textcolor{blue}{#1}}

\newcommand{\oursys}{\texttt{GaussMaster}\xspace}

\setcopyright{none}
\renewcommand\footnotetextcopyrightpermission[1]{} % removes footnote 
\newcommand\vldbdoi{XX.XX/XXX.XX}
\newcommand\vldbpages{XXX-XXX}
\newcommand\vldbvolume{14}
\newcommand\vldbissue{1}
\newcommand\vldbyear{2020}
\newcommand\vldbauthors{\authors}
\newcommand\vldbtitle{\shorttitle} 
\newcommand\vldbavailabilityurl{URL_TO_YOUR_ARTIFACTS}
\newcommand\vldbpagestyle{plain} 

\newcommand{\hi}[1]{\vspace{.25em} \noindent {\bf #1} }
\newcommand{\llm}{\textsc{LLM}\xspace}
\newcommand{\llms}{\textsc{LLMs}\xspace}
\newcommand{\bfit}[1]{\textbf{\textit{#1}}}

\usepackage{bm}

\theoremstyle{definition}

\makeatletter
\newcommand{\removelatexerror}{\let\@latex@error\@gobble}
\makeatother

\begin{document}

%\title{GaussMaster: Practice of Tampering \llm as GaussDB Expert %in Huawei 
%[Industry]}

%\title{\oursys: Practice of LLM-as-a-DBA in GaussDB}
\title{\oursys: LLM-based GaussDB Technician}

%\title{\oursys: An LLM-based Tuning and Diagnosis System} Tuning and Diagnosis 

\title{\oursys: An LLM-based Database Copilot System}

\pagenumbering{arabic}

\author{Wei Zhou$^1$, Ji Sun$^1$, Xuanhe Zhou$^3$, Guoliang Li$^2$, Luyang Liu$^1$, Hao Wu$^1$, Tianyuan Wang$^1$}
\affiliation{%
  \institution{$^{[1]}$ Huawei Technologies Co., Ltd. $^{[2]}$ Tsinghua University $^{[3]}$ Shanghai Jiaotong University}
}
\email{{zhouwei324,sunji11,liuluyang2,wuhao401,wangtianyuan5}@huawei.com;liguoliang@tsinghua.edu.cn;zhouxh@cs.sjtu.edu.cn}

\begin{abstract}
\begin{sloppypar}
% \zxh{rewrite based on intro}
%Database administrators (DBAs) have to handle various routine tasks such as product-specific inquiries, performance troubleshooting, and anomaly diagnosis. Automating some of these tasks can offer significant benefits (e.g., reducing the cost of ownership). However, existing LLM-based database products primarily focus on NL2SQL, and diagnosis-relevant research lacks many necessary functions, such as generating a workload diagnosis report (WDR) with private tools. Thus, we demonstrate \oursys, a system that automates three key phases in a commercial database (GaussDB): \emph{(1) Development phase:} \oursys\ acts as a ``code copilot'', assisting with GaussDB kernel code reviews and test-case generation. \emph{(2) Deployment \& usage phase:} \oursys\ serves as a product-specific ``Q\&A assistant'', leveraging a hybrid knowledge base (e.g., user manuals, FAQs, version-specific details) to provide concise, accurate, and secure responses. \emph{(3) Maintenance phase:} \oursys\ functions as an ``anomaly {resolver}'', orchestrating diagnosis-tree-driven workflows, analyzing real-time system metrics, and delivering precise root-cause identification alongside effective solutions. We have implemented \oursys in real scenarios (e.g., banking), demonstrating notable improvements in question-answering accuracy and overall operational stability.

In the financial industry, data is the lifeblood of operations, and DBAs shoulder significant responsibilities for SQL tuning, database deployment, diagnosis, and service repair. In recent years, both database vendors and customers have increasingly turned to autonomous database platforms in an effort to alleviate the heavy workload of DBAs. However, existing autonomous database platforms are limited in their capabilities, primarily addressing single-point issues such as NL2SQL, anomaly detection, and SQL tuning. Manual intervention remains a necessity for comprehensive database maintenance. \oursys aims to revolutionize this landscape by introducing an LLM-based database copilot system. This innovative solution is designed not only to assist developers in writing efficient SQL queries but also to provide comprehensive care for database services. When database instances exhibit abnormal behavior, \oursys is capable of orchestrating the entire maintenance process automatically. It achieves this by analyzing hundreds of metrics and logs, employing a   Tree-of-thought approach to identify root causes, and invoking appropriate tools to resolve issues. We have successfully implemented \oursys in real-world scenarios, such as the banking industry, where it has achieved zero human intervention for over 34 database maintenance scenarios. In this paper, we present significant improvements in these tasks with code at \blue{\url{https://gitcode.com/opengauss/openGauss-GaussMaster}}.

\end{sloppypar}
\end{abstract}

\maketitle

\iffalse
\pagestyle{\vldbpagestyle}
\begingroup\small\noindent\raggedright\textbf{PVLDB Reference Format:}\\
\vldbauthors. \vldbtitle. PVLDB, \vldbvolume(\vldbissue): \vldbpages, \vldbyear.\\
\href{https://doi.org/\vldbdoi}{doi:\vldbdoi}
\endgroup
\begingroup
\renewcommand\thefootnote{}\footnote{\noindent
This work is licensed under the Creative Commons BY-NC-ND 4.0 International License. Visit \url{https://creativecommons.org/licenses/by-nc-nd/4.0/} to view a copy of this license. For any use beyond those covered by this license, obtain permission by emailing \href{mailto:info@vldb.org}{info@vldb.org}. Copyright is held by the owner/author(s). Publication rights licensed to the VLDB Endowment. \\
\raggedright Proceedings of the VLDB Endowment, Vol. \vldbvolume, No. \vldbissue\ %
ISSN 2150-8097. \\
\href{https://doi.org/\vldbdoi}{doi:\vldbdoi} \\
}\addtocounter{footnote}{-1}\endgroup

\ifdefempty{\vldbavailabilityurl}{}{
\vspace{.3cm}
\begingroup\small\noindent\raggedright\textbf{PVLDB Artifact Availability:}\\
The source code, data, and/or other artifacts have been made available at \url{\vldbavailabilityurl}.
\endgroup
}
\fi

% Gauss LLM-based GaussDB Technician
\vspace{-1em}
\section{Introduction}
\label{sec: intro}

% A single database product provides more than 10,000 pages of developer manuals, administrator guides, O\&M optimization manuals, and fault locating manuals. Common DBAs cannot fully master them. First-line and second-line O\&M personnel spend a lot of time answering customer questions by phone, 90\% of which are basic questions.
% When emergency problems such as inter-product interaction and cross-component problems occur, multi-layer chain and multi-product coordination are often performed to locate the problems. Therefore, the most accurate information cannot be obtained in one step.
% The number of instances in a region is large, and the number of SREs is small. Enterprises at key sites will invest part of the assurance, which brings pressure to enterprise operations.
% The intelligent O\&M platform provides exception analysis and optimization suggestions based on the current system running status and statistics. However, after the database version and service scenario change, real-time adjustment cannot be performed based on different load scenarios, and the generalization capability is weak.

Large language models (LLMs) have shown significant potential to transform the database field~\cite{llm4data, llm4dataDisscuss}.  However, most existing LLM-powered database solutions focus narrowly on natural language to SQL (NL2SQL) tasks~\cite{oracle, amazon, azure, zhou2025cracksql, zhou2025cracksqldemo}, which helps reduce users’ learning curve but does not fundamentally ease the burdens of database technicians. In practice, database professionals must address a wide array of technical issues, ranging from \emph{product-specific inquiries} (often dependent on proprietary knowledge) to \emph{performance troubleshooting} and \emph{anomaly diagnosis} (e.g., precisely identifying which SQL statements spike CPU usage). Although some studies have investigated LLM-based database maintenance~\cite{dbot, pandas, llmdba}, they tend to provide generic solutions and often struggle with tasks like pinpointing the exact slow queries or managing highly specific operational procedures demanded by enterprise-scale environments.

\begin{figure}[!t]
  \centering
  \includegraphics[width=\linewidth]{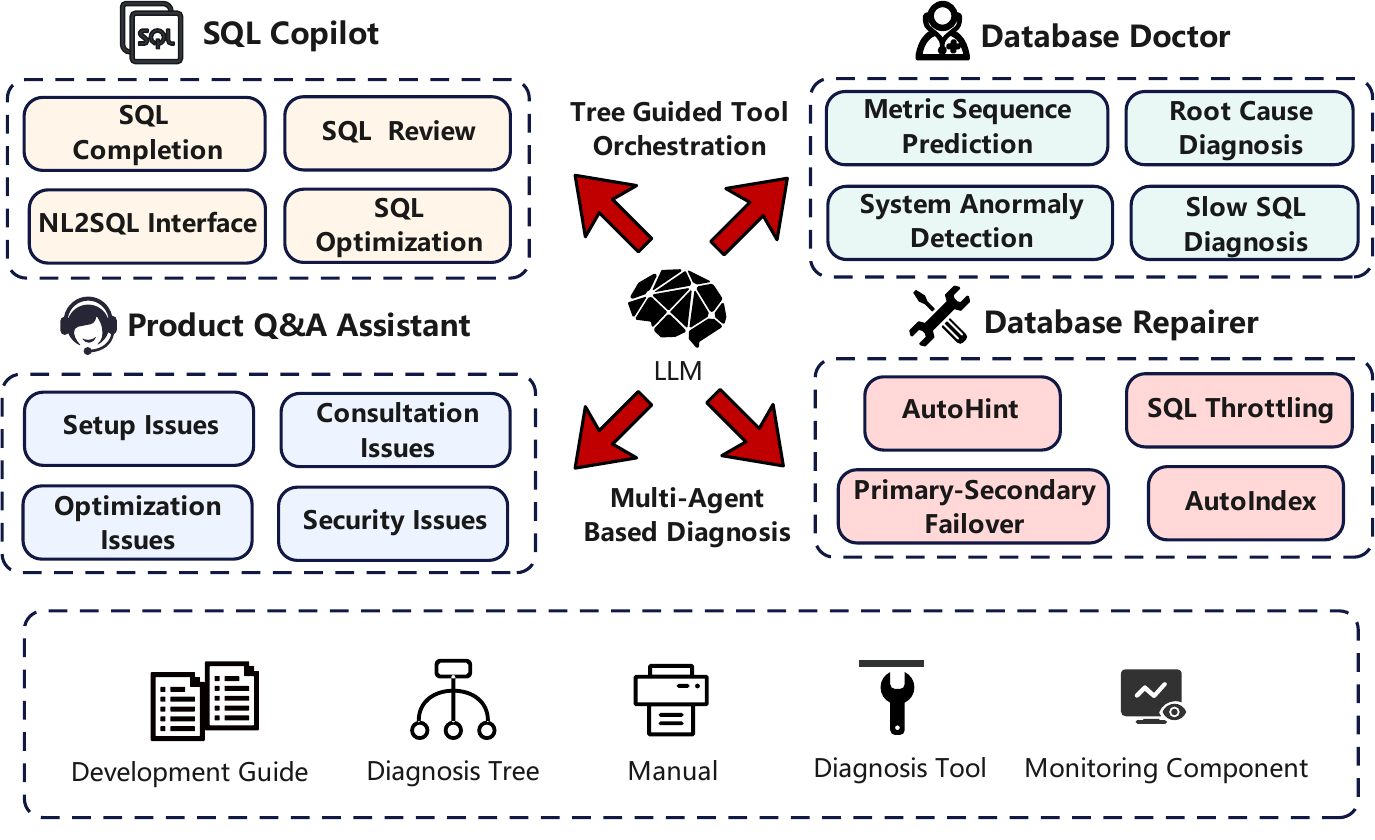}
  \vspace{-1em}
  %\caption{\oursys-GaussDB Lifecycle Technician \zw{bad!}.}
  \caption{\oursys offers comprehensive database maintenance ability across diverse tasks.}
  \label{fig: intro}
  \vspace{-2.5em}
\end{figure}

To bridge these gaps, we demonstrate \oursys, a system that leverages LLMs to function as database technicians (specifically designed for a commercial database named GaussDB) throughout the lifecycle, including stages like database development, deployment, and maintenance (see Figure~\ref{fig: intro}). %As illustrated in Figure~\ref{fig: intro}, \oursys~supports three key phases in industrial GaussDB scenarios: \emph{(1) Development phase:} \oursys~acts as a ``code copilot,'' assisting developers with GaussDB kernel code review, test-case generation \zxh{add some details}. \emph{(2) Deployment \& usage phase:} \oursys~serves as a product-specific Q\&A assistant, tapping into a hybrid knowledge base of user manuals, FAQs, and version-specific details to provide concise, accurate, and safe answers. \emph{(3) Maintenance phase:} \oursys~functions as an ``anomaly doctor'', orchestrating tool-driven diagnosis workflows, analyzing system metrics in real time, and offering precise root-cause identification with prescriptive remedies. %To this end, we propose a demonstration system \oursys, empowering \llm to handle the daily jobs of database technicians in GaussDB. As shown in Figure~\ref{fig: intro}, in the development phase (e.g, \zxh{xxx}), \oursys~serves as a code copilot to help developers write kernel codes better including code review and test case generation. In the deployment \& usage phase (e.g, \zxh{the user needs to deploy GaussDB on a new OS system}), \oursys~acts as a Q\&A assistant for diverse GaussDB questions from users (e.g., {how to enable the AI features in GaussDB?}). In the maintenance phase (e.g., the instance can have various problems like slow responses and crushing), \oursys~serves as an anomaly doctor to pinpoint the root causes of occurring anomaly and make the charged databases well-behaved.
%In summary, \oursys~achieve these functionalities by addressing the following limitations when directly applying \llms in real banking applications. 
% requiring nearly 30\% of the time to handle over 2000 consults.
We address three main challenges when directly applying general-purpose LLMs to the database tasks. First, proprietary GaussDB knowledge is not publicly available, rendering off-the-shelf LLMs incapable of accurately answering many domain-specific questions. Our analysis shows that out of the \bfit{1614 real-world bank questions}, generic LLMs often either fail to provide conclusive answers or incorrectly infer from other database systems (e.g., PostgreSQL rather than GaussDB). Second, high-stakes enterprise environments, such as banking, demand strict correctness and safety, yet LLMs easily produce irrelevant or even hazardous responses. For instance, \llms may hallucinate unsupported commands or carry out risky operations (e.g., unauthorized data access). Finally, LLM-based anomaly diagnosis in GaussDB requires orchestrating multiple private tools in sequence (e.g., slow-query tracing, metric inspections), but naïve LLM-driven pipelines frequently misconfigure tools or invoke them repeatedly, making the diagnostic process unnecessarily long and error-prone.

\iffalse
\begin{sloppypar}
%\noindent \textbf{L1: Lack of specialized GaussDB knowledge}.
\noindent \textbf{L1: {Inefficient Way of Utilizing Heterogeneous GaussDB Knowledge}}.
Based on the analysis over questions for our banking customers, we notice that \llms are inherently unaware of most GaussDB-specific knowledge, which is not publicly available on the online website.
Specifically, out of most of the 1614 questions in real-world bank scenarios, \llms either response they cannot conclude the corresponding answer (e.g., for the question ``\zxh{xxx}'') or make some attempts to infer from other databases like PostgreSQL, but may not work for GaussDB.
For example, for the question ``\zxh{xxx}'', {GPT-4o might conclude that GaussDB cannot generate the WDR report for diagnosis since it is not supported in PostgreSQL}.
\end{sloppypar}
\fi

We address these challenges by integrating (1) a heterogeneous GaussDB knowledge base, (2) an \llm-based Q\&A module, and (3) an \llm-based Diagnosis \& Repair framework into a single end-to-end solution. First, \oursys\ aggregates diverse GaussDB documents and diagnostic tools into a unified knowledge base, covering everything from maintenance FAQs to internal command specifications. Second, a retrieval-augmented generation (RAG) pipeline combines literal and semantic retrieval with multi-stage safety checks to ensure concise, accurate, and risk-free Q\&A responses. Finally, \oursys\ employs an adaptive anomaly diagnosis module. Guided by a series of expert-defined diagnosis trees, this module systematically orchestrates tool usage, fills parameters through a two-step prompt design, and generates an interpretable root-cause report. In real-world banking scenarios (e.g., the Agricultural Bank of China), \oursys\ delivers over 80\% fully correct, safe answers in Q\&A, and achieves over 95\% accuracy in tool orchestration for anomaly diagnosis. %—making it a practical, LLM-empowered technician for the GaussDB lifecycle.

\iffalse
\begin{sloppypar}
To address the above limitations and make \llms become reliable and competitive GaussDB technicians, \oursys makes the following technical contributions:
%we propose a powerful LLM-based system \oursys~based on our practice in banking scenarios. Specifically, we enhance \llm capabilities in terms of three aspects:
\end{sloppypar}

(1) \oursys~involves the building of heterogeneous GaussDB knowledge base \zxh{no techniques!!}, including the specification documents of multiple sources and formats as well as the basic diagnosis tools, covering the maintenance Q\&A and anomaly diagnosis knowledge base.

(2) \oursys~supports an intelligent product-specific Q\&A module with the RAG paradigm to improve answer quality, employing a hybrid knowledge retrieval strategy and further enhanced by a multi-stage safety mechanism \zxh{add details}. 
% zw{rephrase, only the techniques}

(3) \oursys~comprises of an adaptive anomaly diagnosis module, which improves \llm stability by two-step tool orchestration assisted by the well-defined \zxh{diagnosis tree} of expertise diagnosis experience and report generation of root cause prompt templates. 
% \zw{rephrase, only the techniques}
\fi
%presents the overall workflow of two primary modules in \oursys.  For an input natural question from users, \oursys~first conducts intent recognition by decomposing the composite question into several single ones via rule-based syntax parsing and LLM prompting. Then, it assigns the task-specific prompts to the underlying \llms for each module.

\begin{figure}[!t]
  \centering
  \includegraphics[width=\linewidth]{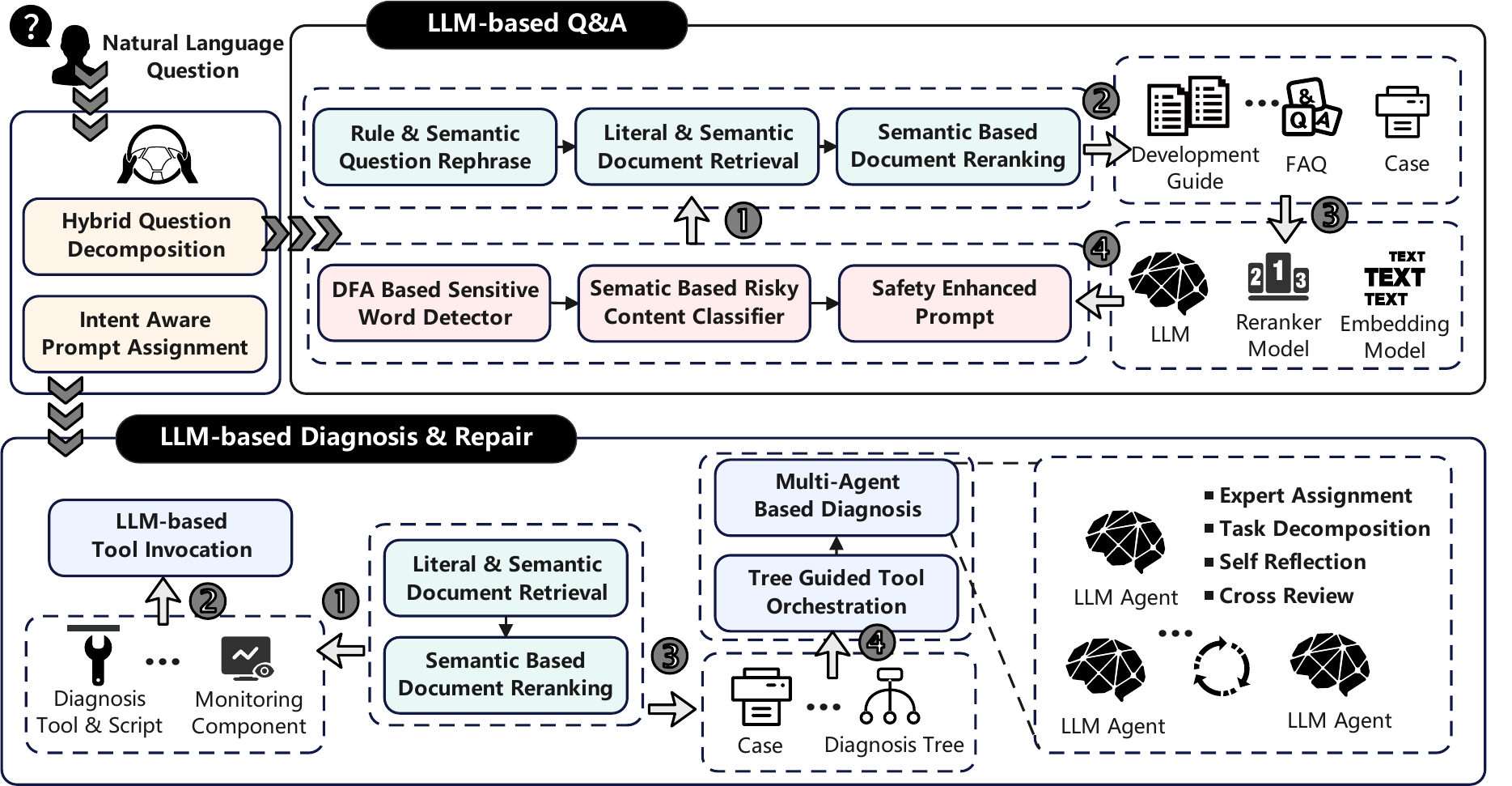}
  \vspace{-1.25em}
  \caption{Overview of \oursys}
  \label{fig: overview}
  \vspace{-1.5em}
\end{figure}

\section{System Implementation}
\label{sec: overview}

\subsection{System Overview}

\oursys~handles diverse database tasks with a unified natural language interface. Figure~\ref{fig: overview} shows the system architecture.

\noindent \textbf{Knowledge Base Preparation.} \oursys~aggregates multi-source GaussDB documents into a unified knowledge base, splitting text by semantic boundaries ({e.g., the code block including GaussDB SQLs}), retaining structural elements ({e.g., the hierarchical titles in markdown files}), and removing duplicates. Each chunk is augmented with version tags and neighboring context. Additionally, \oursys~wraps DBMind’s 25 diagnostic tools under a RESTful interface and codifies expert workflows into anomaly diagnosis trees, improving the troubleshooting stability of \llms.

\noindent \textbf{\llm-based Question Pre-processing.} When a user raises a natural language question, \oursys~first performs two steps of pre-processing: (1) {Hybrid Question Decomposition}: the question is decomposed into finer-grained sub-queries (e.g., smaller semantic chunks or tool-invocation hints) via both rule-based and \llm methods. (2) {Intent-Aware Prompt Assignment}: Based on the question’s content and intent, \oursys~automatically routes {by prompting \llms} the question into either the \emph{\llm-based Q\&A} module {(including the code copilot functionality like code review and optimization)} or the \emph{\llm-based Diagnosis \& Repair} module (if it detects an anomaly-related or troubleshooting intent).

\noindent \textbf{\llm-based Q\&A.} We provide three main functions: (1) Risk checking mechanism with both a sensitive word detector and a semantic-based content classifier. Once the preprocessed questions are considered to be risky, \oursys~returns a pre-defined refusal response to users; (2) Hybrid knowledge retrieval for matching the most relevant information (from sources like the development guide and the FAQ textbook) to answer the question. We fine-tune an embedding model and a re-ranking model over the GaussDB corpus with {106,810} samples by expanding the input question set through rule-based synonym substitution and LLM-based question rephrasing. The retrieved information is utilized to generate high-quality answers;
(3) The generated answer goes through the same risk checking mechanism to ensure safety before responding to users.

\noindent \textbf{\llm-based Diagnosis \& Repair.} We call the module given the description of triggered alerts, which supports four main functions: (1) Identify the correct tools through hybrid retrieval and re-ranking based on user question and tool usage descriptions; (2) Multi-step strategy to fill in the corresponding parameters by analyzing user questions and asking for the complement of absent ones {from users}; (3) Retrieve the relevant historical alarm cases and diagnosis tree (i.e., the sequence of troubleshooting steps by GaussDB experts) for the subsequent diagnosis; (4) Conduct diagnosis-tree-guided orchestration using multiple tools, combined with a multi-agent diagnosis approach that includes expert assignment, task decomposition, and self-reflection~\cite{dbot}.

% \zw{sync with new Fig.}

\iffalse
\noindent \textbf{Workflow}.
As shown in Figure~\ref{fig: overview}, \oursys~is implemented in a three-layer structure.
(1) The \emph{Frontend Console Layer} offers a unified user-friendly web interface to access various services in \oursys;
(2) The \emph{Business Logic Layer} coordinates the invocations of various services through distinct RESTful APIs for different user requests;
(3) The \emph{Computation Layer} implements the inference of the underlying \llm or embedding model, along with the invocations of diagnosis tools in DBMind.
Specifically, an \llm-based Q\&A request from the user interface involves \oursys~server to invoke the embedding model service for input question vectorization, retrieve relevant knowledge materials stored in GaussDB, turn to reranking model service to sort the retrieved documents via calculated importance scores, and rely on \llm service to generate the final answer.
Apart from the aforementioned service invocation, a database fault diagnosis request further needs to invoke the diagnosis tools in DBMind that monitor the status of the GaussDB cluster by various metric collection tools.
\fi

\subsection{Knowledge Base Preparation}
\label{sec: information}

The knowledge base of \oursys~involves multi-source GaussDB information, which is absent in existing \llms.
Currently, there are primarily three types of information.

%  (e.g., development guide and production cases)

\hi{Multi-Source Document Processing.} We process GaussDB documents from multiple sources (e.g., the online community) in four steps. $(1)$ {Type-Specific Parsing.} To maintain the document structure, we implement distinct parsing logic for each document type (e.g., \textsf{``.md''} and \textsf{``.docx''}).  For example, we process the documents (\textsf{``.md''}) considering the multi-section organization characteristics (e.g., retaining the text of markdown syntax for titles of different levels). $(2)$ {Semantic-based Splitting.} Instead of segmenting the parsed documents into a fixed-length chunk, we employ a DNN-based model (supervised by the punctuation of well-organized documents) to ensure paragraphs of the same semantics are within the same chunks with high probability. Besides, we ensure the single table and code snippet are not split apart with varied chunk lengths. $(3)$ {Literal De-duplication.} To mitigate the retrieval introduced by redundant documents, we perform global de-duplication through the hash value of the textual content of the split chunks.

\hi{Document Meta-Info Augmentation.} To complement retrieved documents with absent but relevant ones, we add meta information to each chunk, including the specific database version, the {ids} of previous and next chunks. 

% {(with 277.7k+ code lines)}
\hi{Tool Integration \& Usage Guidance.} $(1)$ {Diagnosis Tool.} We leverage the DBMind tools~\cite{dbmind} to assist \llm in more accurate diagnosis. There are currently {25} tools in total, covering a wide range from metric monitoring, slow query tracing, and optimization solution recommendation with diverse heuristic and learning methods. We deploy each tool in DBMind with a unique RESTful API address to acquire the results in a unified format. $(2)$ {{Anomaly Diagnosis Tree.}} We incorporate human expertise in diagnosis to orchestrate the diagnosis tool invocation pipeline for different database anomalies, which reduces the overhead and improves the accuracy for \llm diagnosis (i.e., pinpoint the correct root cause within a few explorations). %\zw{add statistics}

\subsection{\llm-based Q\&A}
\label{sec: qa}

We employ the following strategies to guarantee the answer quality for strict requirements in scenarios like banking~\cite{autoindex}.

\noindent \textbf{Hybrid Literal and Semantic Retrieval \& Rerank.}
Since a single question might fail to retrieve all the required GaussDB documents, we first expand the questions with rule-based synonym substitutions and semantic-based rephrasing of expressions in the questions via LLM prompting. With the expanded questions, we then perform hybrid retrieval that combines the sparse retrieval algorithm implemented in GaussDB and the dense retrieval with a fine-tuned embedding model over the GaussDB corpus. To handle the potential redundancy in the retrieved documents, we employ a reranker (also fine-tuned over the GaussDB corpus) to calculate the corresponding relevance scores and sort documents in descending order.
In particular, we remove the documents whose scores (computed by the reranker) are below zero to further guarantee the relevance of the retrieved documents. 

\noindent \textbf{Risky Question Control.}
To ensure unsafe response generation is blocked in strict online scenarios, we propose risky question control to prevent risky database operations (e.g., unauthorized access) or biased unfair opinions (e.g., ethnic prejudice).
Once risk is detected in the input question or generated answer, \oursys~returns a response like ``\oursys cannot answer such a question''.

\begin{sloppypar}
\emph{(1) Word Detector \& Content Classifier:}
To enable \oursys~identify both explicit (e.g., specific unsafe commands) and implicit (e.g., risky database operations) questions that may need to be blocked, we employ two respective strategies.
We utilize the deterministic finite automaton (DFA) algorithm as a word detector, where each sensitive word corresponds to a specific path in a trie tree ({20000+} words currently).
We fine-tune a language model (i.e., an XLNet-based model trained with 106,810 questions including risky GaussDB and general unsafe questions~\cite{SafetyPrompt}) to enable it to classify the input question within $500ms$.
\end{sloppypar}

\emph{(2) Safety Enhanced Prompt:} 
We conduct safety-aware prompt engineering by automatically appending safety guidelines to the predefined prompt templates.
Specifically, we prompt \llm to identify unsafe questions, i.e., conducting classification of the input question (assessing whether the question belongs to sensitive topics like data privacy) and composing supplementary guidelines (e.g., judging whether the question is highly relevant to GaussDB).

%from humans with two points: (1) assess whether the question belongs to sensitive topics like data privacy, (2) ensure the generated response is highly relevant to GaussDB} to instruct \llm to identify unsafe questions, strictly adhering to return safe responses based on internal capability.

%  (e.g., ``Please only return GaussDB relevant response strictly adhering to the documents'').

% \textbf{Patch} Apart from the prepared documents in Section~\ref{sec: information}, we maintain an additional FAQ information base with incrementally refined questions and the corresponding answers by humans.
% The FAQ information base serves as a lightweight patch (i.e., without the need for model fine-tuning) for \oursys~to support more inquiries.

\vspace{-.35cm}
\subsection{\llm-based Diagnosis \& Repair}
\label{sec: anomaly}

We employ the following strategies to enable \oursys~perform accurate diagnosis through adaptive tool invocation and analysis.

\noindent \textbf{Diagnosis Information Augmentation.}
To enrich the context of anomaly alerts and make \llm more informed of GaussDB anomalies, we employ a multi-step retrieval and reranking mechanism.
We first acquire relevant development documents and historical anomaly cases to expand the alert descriptions.
Then, the enriched descriptions are utilized to match the tools (offered by DBMind~\cite{dbmind}). % and the diagnosis trees (i.e., tool invocation or trouble shooting steps by human GaussDB experts).

\noindent \textbf{Diagnosis Tree-Guided Tool Orchestration.}
Unlike previous methods, which offer great flexibility for \llm explorations~\cite{dbot}, we make \llm-based diagnosis more deterministic and robust (i.e., alleviate noisy explorations) by leveraging the matched DBA maintenance information. The information is represented in the form of a diagnosis tree, where each tree node corresponds to a specific tool (e.g., \textsf{slow\_sql\_rca()} tool for slow SQL diagnosis).
The tree paths correspond to deterministic tool invocation pipelines to pinpoint the root causes of specific database anomalies (e.g., high CPU usage).

\begin{sloppypar}
\noindent \textbf{Multi-Agent Based Diagnosis \& Repair.} 
To enable \oursys~handle complex GaussDB anomalies involving comprehensive analysis over diverse metrics and systems, we propose agent-based diagnosis with the following steps.
\end{sloppypar}

\begin{figure*}[!t]
  \centering
  \includegraphics[width=\linewidth]{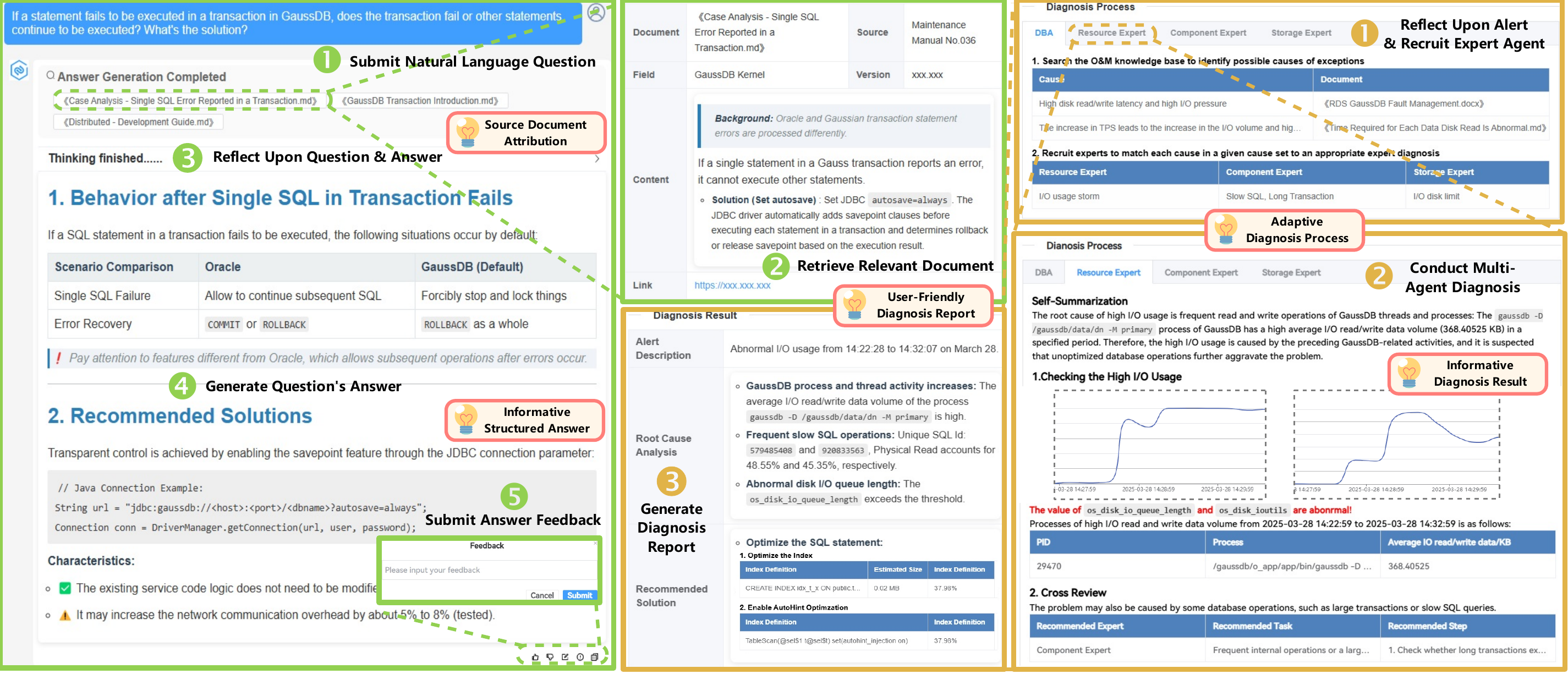}
  % \vspace{-1.75em}
  \caption{Demonstration of \emph{\llm-based Q\&A} and \emph{\llm-based Diagnosis \& Repair} in \oursys.}
  \label{fig: demo}
  % \vspace{-1em}
\end{figure*}

\emph{(1) Expert Assignment \& Task Decomposition:}
To ensure \llm perform more focused diagnosis, we define a hierarchical expert agent structure, where a DBA agent serves as the chief to manage other expert agents.
For each reported alert, the DBA agent conducts an expert assignment to recruit multiple expert agents for specialized diagnosis based on the similarity between the alert and expert descriptions. 
For example, it selects \textsf{Resource Expert} with metric inspection functionality for I/O usage analysis.
Furthermore, it performs task decomposition to provide a series of troubleshooting steps for each expert based on the retrieved documents and the diagnosis trees.

\emph{(2) Self-Reflection \& Cross-Review:}
Unlike other methods that make \llm invoke tools within a single attempt, we enable our expert agent to obtain more accurate tool invocation within a two-step reflection.
Specifically, it first determines the selection of the tool set (i.e., API name in DBMind) based on their functionality descriptions.
Then, the arguments for each selected tool are subsequently extracted based on the alerts.
Next, the tools are invoked and the expert agent reflects upon the returned response to conclude the diagnosis result (e.g., the \textsf{os\_disk\_ioutils} metric is abnormal).
Besides, \oursys~implements cross-review protocol to enable information exchange among different agents during the diagnosis process (e.g., the \textsf{Resource Expert} recommends the \textsf{Component Expert} ignored by the DBA agent for long transaction analysis).

\emph{(3) Result Aggregation \& Report Generation:}
To aggregate the diagnosis results among agents and generate user-friendly reports, we first rephrase the obscure raw results from different tools into predefined templates easier for \llm understanding (e.g., filling the returned metric values in the templates).
Then, we {design prompts for each expert agent to specifically summarize the diagnosis steps with prompts tailored for different root causes} for accurate diagnosis (e.g., the analysis guidelines to emphasize focus on CPU usage).

\vspace{-.35cm}
\section{Application Scenarios}
\label{sec: app}

\subsection{Real World Demonstrations}
\label{sec: demo}
% We demonstrate \oursys~effectiveness in two real-world cases.

We showcase the usage of \oursys in two typical scenarios.

\begin{sloppypar}
\hi{\llm-based Q\&A.} We demonstrate how \oursys~answers NL questions in banking scenario. 
As shown in Figure~\ref{fig: demo}, users first submit their question about GaussDB transactions (\ding{182}).
Then, \oursys~performs the multi-step retrieval and reranking to acquire the information from multi-source documents like transaction descriptions (\ding{183}) and reflects upon how to generate high-quality answers (\ding{184}).
Next, \oursys~generates answers ``1. Behavior after single...'' rendered in markdown formats in a streaming manner (\ding{185}).
Last, users can provide feedback about the answers by clicking the button \emph{``report a missing solution''} (\ding{186}). 
% The above steps can be repeatedly conducted in multiple rounds, where users can perform continuous inquiries and \oursys replies based on the dialogue history.
\end{sloppypar}

\iffalse
\noindent \textbf{Takeaway.}
\oursys~supports \emph{Source Document Attribution} by providing the details of the relevant documents including their meta information (e.g., the source and the field) and original content.
Furthermore, \oursys~generates \emph{Informative Structured Answer} with two listed points for each inquiry of the input GaussDB question, illustrative code snippets, and informative table summary.
\fi

\begin{sloppypar}
\hi{\llm-based Diagnosis \& Repair.} 
% due to execution of a stored procedure for inserting large fields and a SQL with heavy scans over large tables
We demonstrate how \oursys~pinpoints the root cause of an abnormally high I/O usage in a real-world bank scenario without any human intervention.
As shown in Figure~\ref{fig: demo}, for a reported alert ``\textsf{Abnormal I/O Usage}'', DBA agent first reflects upon the details (e.g., retrieved similar historical anomaly cases) and recruits other expert agents (e.g., \textsf{Resource Expert Agent}) for in-depth diagnosis (\ding{182}).
Then, each recruited expert agent (e.g., \textsf{Resource Expert Agent}) performs diagnosis over a series of tasks (e.g., \textsf{Checking the High I/O Usage}), conducts cross review among agents (e.g., turn to \textsf{Component Expert} for slow SQL analysis), and presents self-summarization over the whole diagnosis process (\ding{183}).
Last, \oursys~presents user-friendly report with comprehensive information (\ding{184}).
\end{sloppypar}

\iffalse
\noindent \textbf{Takeaway.}
\oursys~undergoes \emph{Adaptive Diagnosis Process} with anomaly-specific expert agents recruited by the DBA agent based on historical cases from hybrid retrieval and reranking.
Furthermore, \oursys~provides \emph{Informative Diagnosis Result} with comprehensive textual analysis (e.g., high I/O usage of GaussDB process) as well as illustrative metric charts (e.g., abnormal \textsf{os\_disks\_utils} increasing) and table (e.g., GaussDB process with PID 29470 occupies high I/O usage).
Most importantly, \oursys~presents \emph{User-Friendly Diagnosis Report} with detailed root cause analysis (e.g., frequent slow SQL operations with ID 579485408 and 920833563) and practical recommended solutions (e.g., optimize SQL by index).
\fi

\begin{figure}[!t]
    \centering
    \begin{subfigure}[b]{0.19\textwidth}
        \includegraphics[width=\textwidth]{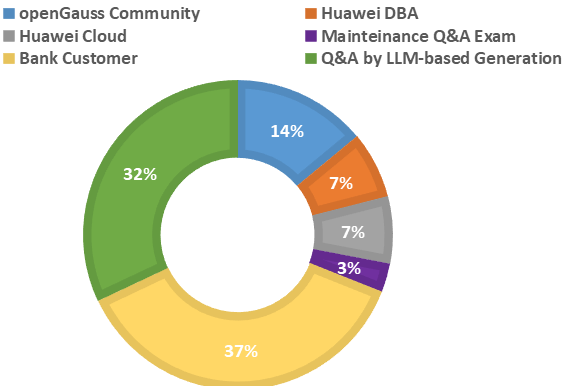}
        \caption{Question Distribution}
        \label{fig: inquiry_acc_p1}
    \end{subfigure}
    % \hfill % 这里添加了一个水平填充以增加间距
    \begin{subfigure}[b]{0.25\textwidth}
        \includegraphics[width=\textwidth]{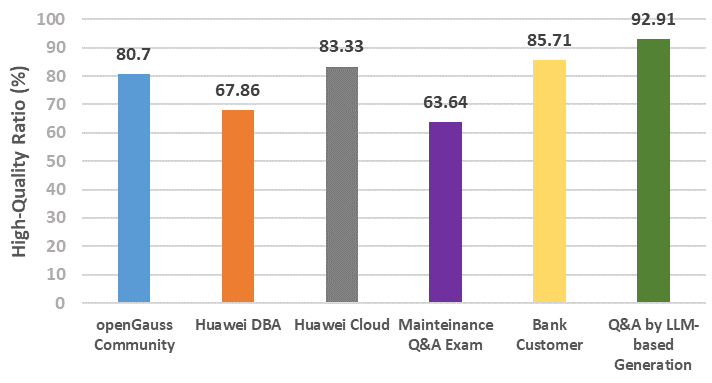}
        \caption{Answer Quality Distribution}
        \label{fig: inquiry_acc_p2}
    \end{subfigure}
    % \vspace{-1em}
    \caption{Answer Quality of Testing Questions.}
    \label{fig: inquiry_acc}
    % \vspace{-1em}
\end{figure}

% \vspace{-.8cm}
\subsection{Experiment Results}
\label{sec: exp}
%\blue{find a scenario which contains a complete pipeline,not just accuracy}
We assess the performance of \oursys, i.e., (1) the answer quality of \emph{\llm-based Q\&A} over diverse GaussDB questions and (2) the tool invocation accuracy of \emph{\llm-based Diagnosis \& Repair} over different database anomalies. {The underlying \llm is Pangu-38B (a commercial model in Huawei) by default.}

\begin{sloppypar}
\noindent \textbf{Answer Quality.}
We adopt a hybrid evaluation strategy where both the human and \llms serve as the judges. And we simultaneously utilize the user questions, standard answers, and \oursys answers to assist the decisions, i.e., whether answers generated by \oursys~meet the standard ones in terms of three aspects (i.e., relevance, accuracy, and safety) in Section~\ref{sec: qa}.

As shown in Figure~\ref{fig: inquiry_acc}, we observe \emph{\oursys~achieves high answer quality across 400+ real-world questions}.
Specifically, the ratio of high-quality answers arrives at {85.23\%} on average.
The underlying reason is that \oursys~informs \llm with the necessary document information (via multi-step retrieval) to improve the answer relevance and accuracy. Moreover, it prevents \llm from generating risky responses via the safety mechanism ({Section~\ref{sec: qa}}).
\end{sloppypar}

\begin{sloppypar}
\noindent \textbf{Tool Invocation Accuracy.}
We assess whether \oursys~can identify the correct diagnosis tool as well as the corresponding arguments~\cite{dbmind}, providing the illustrations of different database anomalies.
We observe \emph{\oursys~can effectively invoke the diagnosis tool with a 95\%+ tool selection accuracy and 99\%+ tool parameter filling accuracy on average}. 
The underlying reason is that \oursys decomposes tool invocation into steps, alleviating the difficulty of a one-time strategy and facilitating the required tool orchestration to get the final comprehensive report.
\end{sloppypar}

\iffalse
\begin{table}[!t]
\caption{\oursys~Tool Invocation Accuracy.}
% \vspace{-.35cm}
\label{tab: tool_acc}
\resizebox{0.7\linewidth}{!}{
\begin{tabular}{lcc}
\hline
\multicolumn{1}{c}{\textbf{Tool}} & \textbf{Tool Selection} & \textbf{Parameter Filling} \\ \hline
\textbf{(w/o params)}             & 0.99                    & 1.00                       \\
\textbf{(w/ params)}              & 0.99                    & 0.95                       \\ \hline
\end{tabular}
}
\end{table}
\fi

%\clearpage
%\newpage

\bibliographystyle{ACM-Reference-Format}
\bibliography{software}

\end{document}